\documentclass[9pt,twocolumn,twoside]{osajnl}

\journal{ol} 

\setboolean{shortarticle}{true}
\usepackage{multicol}
\title{Synthetic topological insulator with periodically modulated effective gaugue fields} 

\author[1,2]{Zengrun Wen}
\author[1,2,*]{Baole Lu}
\author[3]{Kaiwen Ji}
\author[1,2]{Kaile Wang}
\author[1,2]{Haowei Chen}
\author[3]{Xinyuan Qi}
\author[1,2,*]{Jintao Bai}

\affil[1]{State Key Laboratory of Photoelectric Technology and Functional Materials, International Collaborative Center on Photoelectric Technology and Nano Functional Materials, Institute of Photonics and Photon-technology, Northwest University, Shaanxi, Xi'an 710069, China}
\affil[2]{Shaanxi Engineering Technology Research Center for Solid State Lasers and Application, Provincial Key Laboratory of Photo-electronic Technology, Northwest University, Shaanxi, Xi'an 710079, China}
\affil[3]{School of Physics, Northwest University, Xi'an 710069, China}

\affil[*]{Corresponding author: lubaole1123@163.com; baijt@nwu.edu.cn}

\begin{abstract}
We study  both theoretically and numerically the  topological edge states in synthetic photonic lattice with finitely periodic gauge potentials. The effective gauge fields are implemented by tailoring the phase alternatively and periodically, which finally results in symmetric total reflection at two boundaries of the one-dimensional synthetic lattice. Further tuning the nearest-neighbor coupling anisotropically, topological edge states occur at the two boundaries. Our work provides a new way to study the topological physics of one-dimensional coupled waveguide arrays with synthetic photonic lattice.
\end{abstract}

\setboolean{displaycopyright}{true}

\begin{document}
	
	\maketitle
	Topological photonics, a rapidly developing research area aiming to realize photonic analogy of quantum Hall effects and topological insulators, leads to intriguing phenomena such as robust unidirectional propagation of light and localized edge state in photonic systems~\cite{RevModPhys.91.015006}. Significantly, various topological systems support robust edge states which are capable of against impurities and imperfections~\cite{wuaom2017}, making it possible to exploit low-loss transmission devices and photonic circuitry immune to disorder. In one dimensional systems, topological edge state can be endowed based on chiral symmetry~\cite{ryunjp2010}. As an example, Su-Schrieffer-Heeger (SSH) model in photonics was initially realized in superlattice~\cite{Malkova:09}, where the localization at the edge is quite different from the one in trivial defect configuration~\cite{prlblanco2016}. Recently, analogy of SSH model has been a general solution to realize topological edge state in various areas in photonics comprising photonic crystals, plasmonic waveguide arrays, coupled optical waveguides and micro-cavity array lasers and so on~\cite{nckeil2013,prbbleckmann2017,pranaz2018,prlpartp2018,longhiadp2018,olji2020}. 
	
	Synthetic photonic lattices (SPLs), a new type of structure comprising of two mutually coupled fiber loops with slightly length difference was reported firstly in 2011~\cite{RegensburgerParity}. Such a structure shares similar principle as the Galton board, which provides a versatile fundamental platform for the understanding of quantum walks and pulse propagation dynamics in discretized periodic waveguide arrays~\cite{prlregensburger2011}. As one of the most typical applications, SPLs has been employed as the one-dimensional waveguides without actually etching the waveguides~\cite{vatniksr2017}. By introducing the modulation from  amplitude or phase modulators driven by certain signals, equivalent potentials of SPLs can be achieved and tuned, which results in various linear or nonlinear waveguide arrays~\cite{WimmerObservation}, as well as parity-time (PT) symmetric system~\cite{pramiri2012}. Based on these, a series of optical propagation phenomena have all been realized or observed in SPLs, including time-reversed light propagation~\cite{RN4}, Talbot effect~\cite{prawang2018}, Kapitza light~\cite{olMuniz2019} and optical diametric drive acceleration through action-reaction symmetry breaking in passive configurations~\cite{WimmerOptical}, Bloch oscillation with periodic outburst of radiation~\cite{WimmerObservationsr} and discrete solitons in PT-symmetric configurations~\cite{WimmerObservation}. Besides, SPLs can also be used to achieve light localization, such as Anderson localizations by inducing random phase~\cite{vatniksr2017} or random coupling coefficients~\cite{VatnikAnderson}; Recently,  defect and gauge field surface were also employed to  obtain the light  localization~\cite{Regensburgerprl2013,Pankov2019Observation}. As a type of robust state immune to the ambient perturbations,  topological localization in  SPL with nonlinearity and non-Hermiticity were investigated very recently~\cite{Bisianovpra,Weidemannsci}. However, the studies on the topological phenomena with periodically modulated effective gaugue fields are still absent.
	
	In this Letter, a SPL with periodic gauge field is constructed with two fiber loops through controlling the phase accumulations and the coupling coefficient of the coupler. Considering the time boundary condition, impulse evolution equations were deduced analytically to investigate the band structure and the impulse responses of the SPL. Further, winding number was employed to study the topological invariant and topological edge modes in the SPL. Numerical simulations of impulse transmissions at the two edges were demonstrated correspondingly. 

\begin{figure}[htbp]
		\centering
		\fbox{\includegraphics[width=\linewidth]{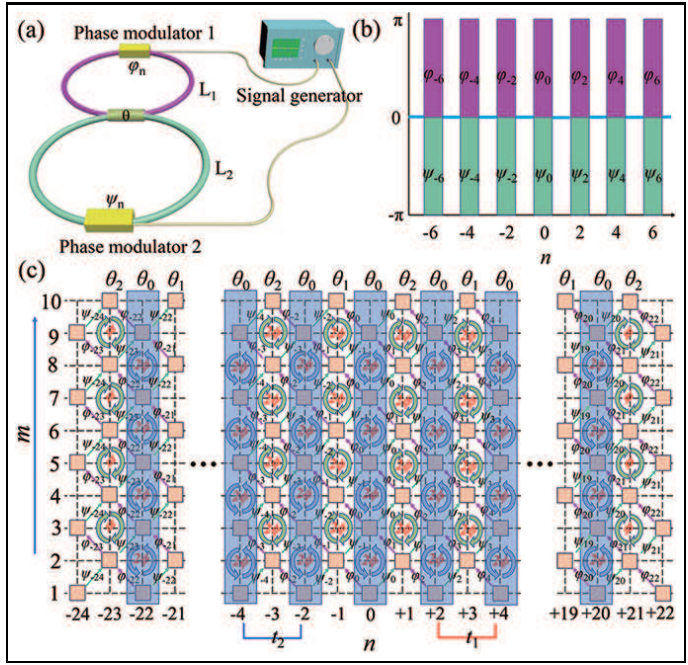}}
		\caption{Synthetic photonic lattice. (a) Schematic of the SPL realized by two mutually coupled fiber loops; (b) phases values $\psi_n$ and $\varphi_n$ controlled by phase modulators; (c) equivalent SPL structure in the presence of gauge fields.}
		\label{fig1}
	\end{figure}

	We constructed a SPL based on a two-fiber-loop system. As pictorially shown in Fig.~\ref{fig1}(a), two fiber loops, each comprising a phase modulator, are connected by a coupler; the two phase modulators subjected to an external signal generator provide time-relevant phase change that forms phase potentials (waveguide pattern) in the SPL. To introduce periodic gauge fields, the added phases display an alternate configuration in transverse direction to avoide the phase overlap between two adjacent gauge fields, as described in Fig.~\ref{fig1}(b). The phase values in short and long loops are limited within $-\pi\sim0$ and $0\sim\pi$, respectively. According to the equivalent principle between fiber loops and SPL~\cite{RegensburgerParity}, our translated model is displayed in Fig.~\ref{fig1}(c). The evolution of the amplitudes in such a SPL can be described by the coupled iterative equation
	\begin{equation}
	\begin{split}
	u_n^{m+1}=e^{i\varphi_n}[\cos\theta_{n+1}u_{n+1}^m+i\sin\theta_{n+1}v_{n+1}^m],\\
	v_n^{m+1}=e^{i\psi_n}[\cos\theta_{n-1}v_{n-1}^m+i\sin\theta_{n-1}u_{n-1}^m],
	\end{split}\label{eq1}
	\end{equation}
	where $\theta$, $m$ and $n$ denote the coupling ratio of the coupler, discrete time coordinate and position coordinate, respectively. The variates $u_n^m$ and $v_n^m$ are light wave amplitudes in short and long fiber loops, respectively. Based on the periodic gauge field configuration, the phase circumfluence is $\varphi_n+\psi_{n-1}-\varphi_{n+1}-\psi_n$. For simplicity, we define $\varphi_n=-\psi_n=\phi$ when $n$ is an even integer and $\varphi_n=-\psi_n=0$ for otherwise $n$. Then the counterclockwise phase circumfluence exists at even position sites (with blue boxes) and clockwise counterpart is induced at odd position sites correspondingly, as depicted in Fig.~\ref{fig1}(c). The coupling coefficient of the coupler at even position sites is represented as $\theta_0$, and $\theta_1$ and $\theta_2$ alternated at odd sites along the position axis. 
	
	When the SPL encompasses gauge fields with finite period $P=22$ (left edge: $n=-22$; right edge: $n=20$), there are phase accumulations in the region of  $-23\leq n \leq 21$ and the clockwise circumflux of $\phi$ are introduced at the adjacent sites outside the two edges. While no phase  accumulation exists at otherwise position sites, namely, the empty lattice, where the light propagates ballistically with rapid divergent speed~\cite{WimmerObservation}. Therefore, it can be considered that there is no waveguide structure in empty lattice area ($n<-23$ and $n>21$), where the optical intensities can be ignored. With the plane wave trial solutions $u_n^m=U_n\exp(-iqn-i\beta m)$, $v_n^m=V_n\exp(-iqn-i\beta m)$, the propagation constant of the position sites with phase circumfluence can be calculated as
	\begin{equation}
	\begin{split}
	\beta_{q,\pm}&=\frac{1}{2}\arccos\{\pm\frac{1}{2}[-\cos\phi\sin\theta_0(\sin\theta_1+\sin\theta_2)\\
	&-\left(2\cos^2(\theta_0)(1+\cos\theta_1\cos\theta_2\cos(q+2\phi)-\sin\theta_1\sin\theta_2)\right.\\
	&\left.+\cos^2\phi\sin^2\theta_0(\sin\theta_1-\sin\theta_2)^2\right)^{\frac{1}{2}}]\}
	\end{split}\label{eq2}
	\end{equation}
	where $q=2l\pi/(2P+1), (l=1,2,...,P)$ is the Bloch wavenumber (it is quantized due to the finite periods of gauge fields). The eigenvalue of the SPL is $\lambda_{n,\pm}=e^{-i\beta_{n,\pm}}, (\beta_n=\beta_{q-2P})$~\cite{Pankovoe}. Obviously, the light transport behavior can be controlled via modulating the values of $\theta_0$, $\theta_1$, $\theta_2$, and $\phi$.
	\begin{figure}[htbp]
		\centering
		\fbox{\includegraphics[width=\linewidth]{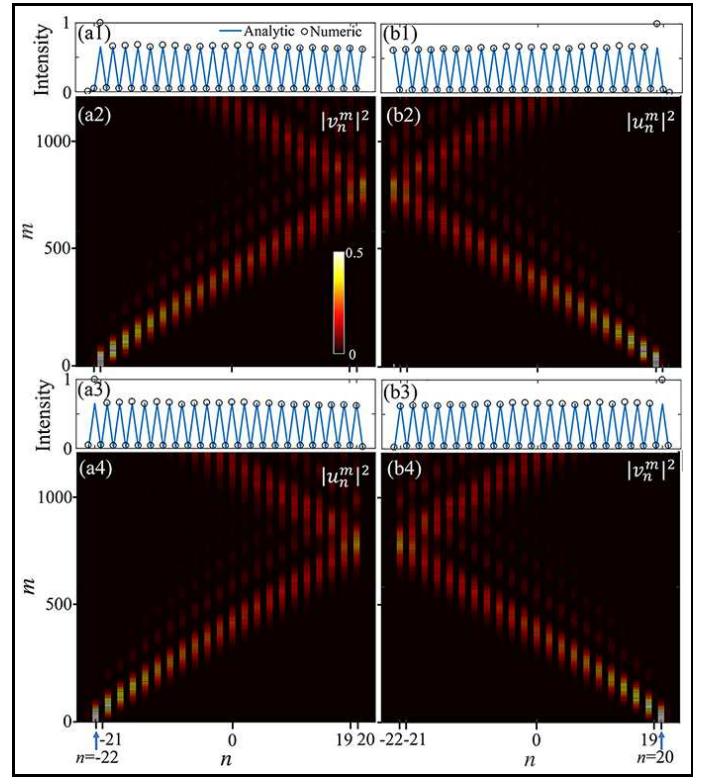}}
		\caption{Impulse responses of the SPL with finite periodic gauge fields. (a1), (a2) $|v_n^m|^2$ and (a3), (a4) $|u_n^m|^2$ with the excitation site ${\it{n}}=-22$; (b1), (b2) $|u_n^m|^2$ and (b3), (b4) $|v_n^m|^2$ with the excitation site ${\it{n}}=20$. In all the cases, $\theta_0$, $\theta_1$ and $\theta_2$ are equal to $0.42\pi$.}
		\label{fig2}
	\end{figure}

	In the previous literature, both identical and different intensity distributions in the two fiber loops in the presence of different phase configurations have been exhibited while there were no detailed explanation~\cite{pramiri2012}. Here, the impulse response principles at the two edges and in the two loops will be investigated theoretically and numerically in our SPL. Substituting the ansatz solutions to Eq. (\ref{eq1}), optical amplitudes in the two fiber loops can be separately expressed, which read
	\begin{equation}
	\begin{split}
	U_{n+1}&=\kappa_{1,n}U_n+\kappa_{2,n}U_{n-1},\\
	V_{n+1}&=\kappa_{1,n}'V_n+\kappa_{2,n}'V_{n-1},
	\end{split}\label{eq3}
	\end{equation}
	where the coefficients are $\kappa_{1,n}=\frac{1}{\cos\theta_{n+1}}(\frac{\lambda_{n,+}}{e^{i\varphi_n}}+\frac{e^{i\psi_{n+1}}\sin\theta_{n+1}}{\lambda_{n,+}\sin\theta_n})$, $\kappa_{2,n}=-\frac{e^{i\psi_{n+1}}\tan\theta_{n+1}}{e^{i\varphi_{n-1}}\tan{\theta_n}}$, $\kappa_{1,n}'=\frac{1}{\cos\theta_n}(\frac{e^{i\psi_{n+1}}\sin\theta_n\lambda_{n,-}}{e^{i(\varphi_{n-1}+\psi_n)}\sin\theta_{n-1}}+\frac{e^{i\psi_{n+1}}}{\lambda_{n,-}})$ and $\kappa_{2,n}'=-\frac{e^{i\psi_{n+1}}\tan\theta_n}{e^{i\varphi_{n-1}}\tan{\theta_{n-1}}}$. According to the boundary condition of the SPL, only the optical amplitudes in the region of $-23\leq n \leq21$ are considered. Thus, it is reasonable to disregard the coefficient in the empty lattice area. Besides, based on Eq.~(\ref{eq1}), the initial inject sites of short loop $u_n^m$ and long loop $v_n^m$ correspond to those of $v_{n+1}^{m+1}$ and $u_{n-1}^{m+1}$ in the other loops, respectively. Therefore, there is always one position site delay of $V_n^m$ in comparison with $U_n^m$~\cite{pramiri2012}.
	When a point source is input into the short loop and the inject site is located at the left edge ($n=-22$),  the optical amplitude as a function of $n$ is derived as
    \begin{equation}
    \begin{split}
    &U_n=\alpha_nU_{-23}, \quad -23\leq n \leq 20\\
    &V_n=\delta_nV_{-22}, \quad\; -22\leq n \leq 21
    \end{split}\label{eq4}
    \end{equation}
    where the coefficients are deduced as  $\alpha_n=\kappa_{1,n-1}\alpha_{n-1}+\kappa_{2,n-1}\alpha_{n-2}$ with initial values of $\alpha_{-23}=1$ and $\alpha_{-22}=\kappa_{1,-23}$, and $\delta_n=\kappa_{1,n-1}'\delta_{n-1}+\kappa_{2,n-1}'\delta_{n-2}$ with $\delta_{-22}=1$ and $\delta_{-21}=\kappa_{1,-22}'$. Clearly, there are identical recurrence and initial condition between $|\alpha_n|$ and $|\delta_{n+1}|$, meaning the intensity distribution of $U_n$ is similar to that of $V_{n+1}$.
    For the simplest case of $\theta_0=\theta_1=\theta_2$, the maximum intensity contrast between two adjacent position sites corresponds to the maximum value of $||\alpha_n|-|\alpha_{n-1}||$ and $\phi=\pi/2$ was obtained, leading to light amplification and attenuation alternately along transverse direction $n$. Under this circumstance, the light intensities in the two loops were numerically calculated based on Eq.~(\ref{eq4}), and the numerical results were obtained by solving Eq.~(\ref{eq1}) for $m=5000$. Both the results are depicted in Figs.~\ref{fig2}(a1) [long loop] and (a3) [short loop], which distribute alternately among high and low values. Besides, the numerical calculation is coincident with the analytic results except intensity difference at the initial inject sites. The discrepancy originates from a part of light energy spreading into the empty lattice area that is disregarded in theoretical model. Corresponding to the optical intensities, the light propagation dynamics in the two loops were numerically simulated, as depicted in Figs.~\ref{fig2}(a2) and (a4). The light propagates rightward into the bulk fastly from the inject site with an alternating bright-dark distribution and a bit chromatic dispersion, which can be in analogy with the transmission features of one-dimensional discrete waveguides. After encountering the right edge of gauge fields, the light is reflected and transport towards left. Consistent with theoretical prediction, the optical response $|v_n^m|^2$ is highly similar to $|u_n^m|^2$, merely with one position site delay. 
    When the light is injected into the long loop from the right edge ($n=20$) of the gauge fields, the optical amplitudes are expressed as
	\begin{equation}
	\begin{split}
	&U_n=\alpha_n'U_{20}, \quad  -23\leq n \leq 20\\
	&V_n=\delta_n'V_{21}, \quad\; -22\leq n \leq 21
	\end{split}
	\label{eq5}
	\end{equation}
	where the coefficients are $\alpha_n'=\frac{1}{\kappa_{2,n+1}}(-\kappa_{1,n+1}\alpha_{n+1}'+\alpha_{n+2}')$ with initial values of  $\alpha_{20}'=1$ and $\alpha_{19}'=-\kappa_{1,20}/\kappa_{2,20}$, and  $\delta_n'=\frac{1}{\kappa_{2,n+1}'}(-\kappa_{1,n+1}' \delta_{n+1}'+\delta_{n+2}')$ with the initial values of $\delta_{21}'=1$ and $\delta_{20}'=-\kappa_{1,21}'/\kappa_{2,21}'$, respectively. Obviously, the recurrence form and initial condition of $\alpha_n'$ and $\delta_n'$ are identically with $\delta_n$ and $\alpha_n$, respectively. Therefore, the impulse responses in the case of right edge injection should be closely symmetric with those in left edge excitation situation, which are verified both in analytic calculation and numerical simulation [Figs.~\ref{fig2}(b1-b4)]. 

\begin{figure}[htbp] 		
		\centering 		
		\fbox{\includegraphics[width=\linewidth]{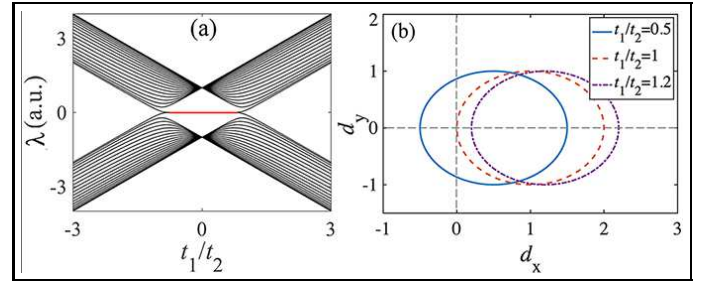}} 		
		\caption{Topological character of the SPL. (a) Numerical band structures as $P=22$; (b) vector $\boldsymbol{d}(k)$ versus different coupling coefficients.} 		
		\label{fig3} 	
\end{figure}

	According to the light propagating principle of the SPL with finite periods of gauge fields, the bright and dark position sites can be regarded as the waveguide and coupling units, respectively. Thus, Eq.~(\ref{eq3}) can be further derived as the following coupled equations through iterative calculation and considering the assumed plane wave solutions
	\begin{equation}
	\begin{split}
	u_n^{m+1}=t_1u_{n-2}^m+t_2u_{n+2}^m, \mod(n,4)=0\\
	u_n^{m+1}=t_2u_{n-2}^m+t_1u_{n+2}^m, \mod(n,4)=2
	\end{split}\label{eq6}
	\end{equation} 
	where $t_1=\Lambda/(\kappa_{1,n}'\kappa_{1,n+1}')$ and $t_2=\Lambda/(\kappa_{1,n},\kappa_{1,n-1})$ with a common factor of $\Lambda=\lambda/[1+\kappa_{2,n+1}/(\kappa_{1,n}\kappa_{1,n+1})+\kappa_{2,n}/(\kappa_{1,n}\kappa_{1,n-1})]$ are the coupling coefficients corresponding to the position sites of $\theta_1$ and $\theta_2$, which are marked in Fig.~\ref{fig1}(c). The above equations are simple enough to calculte the bandstructure. Obviously, the SPL is transformed into a SSH model with a tight-binding chain and alternating coupling constants $t_1$ (for intracell) and $t_2$ (for intercell). The amplitude $V_n$ in the long loop is not proposed due to the similar propagation behavior in the two fiber loops. One can obtain the form of $V_n$ similar to Eq. (\ref{eq6}). Figure~\ref{fig3} depicts the numerical calculated band structures as a function of $t_1/t_2$ according to Eq. (\ref{eq6}), which clearly shows the two bands originating from the upper and lower energy bands degenerate and connect together when $|t_1/t_2|<1$ is satisfied. Besides, the vector $\boldsymbol{d}(k)$ [$d_x=t_1+t_2\cos(2k)$, $d_y=t_2\sin(2k)$, where $k$ is the wavevector of plane beam] in Fig.~\ref{fig3}(b) clearly shows the topological invariant of the SPL. When $|t_1/t_2|<1$ is satisfied, the winding number of SPL is $1$,  indicating there is topological edge state at the edge. Inversely, the SPL is topological trivial and  no topological edge state exists if there is $|t_1/t_2|>1$~\cite{ssh}. The relation of $t_1$ and $t_2$ are expressed as $|t_1/t_2|=|\frac{\cos\theta_1(\sin^2\theta_2+\sin^2\theta_0)}{\cos\theta_2(\sin^2\theta_1+\sin^2\theta_0)}|.$
	Correspondingly, the topological nontrivial condition is translated to $\theta_2<\theta_1<\pi-\theta_2$ in the range of $0<\theta_2<\pi/2$ and $\pi-\theta_2<\theta_1<\theta_2$ in the range of $\pi/2<\theta_2<\pi$, respectively. 
	\begin{figure}[htbp] 		
		\centering 		
		\fbox{\includegraphics[width=\linewidth]{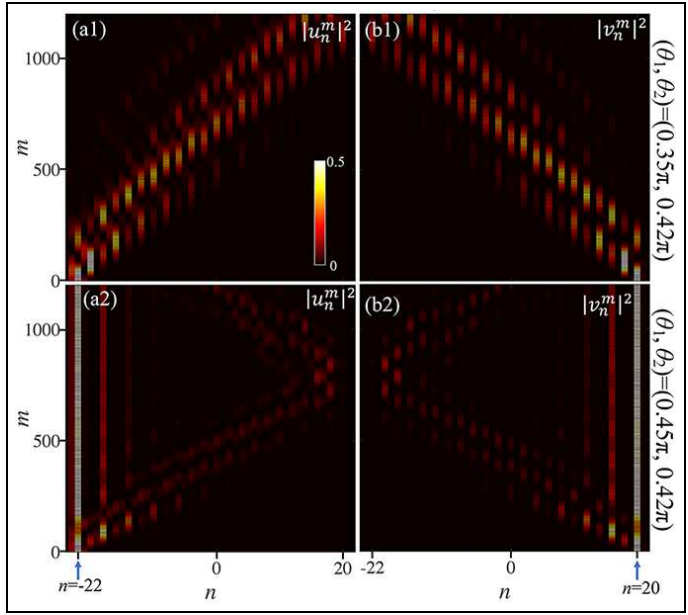}} 		
		\caption{Propagation responses along the edge of the gauge fields: (a1)  and (a2) $|u_n^m|^2$ with the excitation site ${\it{n}}=-22$; (b1) and (b2) $|v_n^m|^2$ with the excitation site ${\it{n}}=20$. In all cases, $\theta_0$ is equal to $0.42\pi$.} 		
		\label{fig4} 	
	\end{figure}

	Figure~\ref{fig4} shows the numerical simulations of impulse responses in our SPL with different values of $\theta_1$ when $\theta_2$ is fixed to $0.42\pi$. When the leftmost position site ($n=-22$) with gauge fields is excited, the light propagates into the bulk if $\theta_1$ equals to $0.35\pi$. While the majority of  light intensity localized at the edge once the coefficient $\theta_1$ satisfies topological condition [see Figs.~\ref{fig4}(a1) and (a2)]. Based on the propagating characteristic of the SPL, the corresponding light dynamics of $|v_n^m|^2$ is similar to Figs.~\ref{fig4}(a1) and (a2) but with rightward translation of one position site. If the light is input at the right edge, the propagation trajectories of $|v_n^m|^2$ are also adhering to the topological condition and almost symmetric with those corresponding to left edge input [see Figs.~\ref{fig4}(b1) and (b2)]. Identically, the light response of $|u_n^m|^2$ is similar to Figs.~\ref{fig4}(b1) and (b2) but with leftward translation of one position site. Apparently, the simulated propagation behaviors are coincident with the theoretical analysis. 
	
	We have achieved a topological insulator in one-dimensional synthetic photonic lattice with finite gauge potentials and anisotropic coupling constants. The finite gauge potentials are realized by tuning the phase modulators alternatively and periodically, which result in the impulse propagation very similar to those in one-dimensional photonic lattices. Further tuning the coupling constants between two nearest lattice sites will lead to robust topological edge states both at the left and right interfaces. The numerical results are in good agreement with our theories. Our work builds a bridge between the researches in topological photonic lattices and the synthetic mesh lattice and may have potential applications in light pulse reshaping in coupled fiber ring systems.
	
	\section*{Acknowledgement}
	This work is supported by National Nature Science Foundation of China (61905193); National Key R\&D Program of China (2017YFB0405102); Key Laboratory of Photoelectron of Education Committee Shaanxi Province of China (18JS113); Open Research Fund of State Key Laboratory of Transient Optics and Photonics (SKLST201805); Northwest University Innovation Fund for Postgraduate Student (YZZ17099).
	
	\section*{Disclosures} The authors declare no conflicts of interest.
	\bibliography{sample}
	
	\bibliographyfullrefs{sample}
	
%
	
	\ifthenelse{\equal{\journalref}{osajnl}}{%
		\section*{Author Biographies}
		\begingroup
		\setlength\intextsep{0pt}
		\begin{minipage}[t][6.3cm][t]{1.0\textwidth} 
			\begin{wrapfigure}{L}{0.25\textwidth}
				\includegraphics[width=0.25\textwidth]{john_smith.eps}
			\end{wrapfigure}
			\noindent
			{\bfseries John Smith} received his BSc (Mathematics) in 2000 from The University of Maryland. His research interests include lasers and optics.
		\end{minipage}
		\begin{minipage}{1.0\textwidth}
			\begin{wrapfigure}{L}{0.25\textwidth}
				\includegraphics[width=0.25\textwidth]{alice_smith.eps}
			\end{wrapfigure}
			\noindent
			{\bfseries Alice Smith} also received her BSc (Mathematics) in 2000 from The University of Maryland. Her research interests also include lasers and optics.
		\end{minipage}
		\endgroup
	}{}

\end{document}